\documentclass[12pt]{iopart}
\usepackage{amsfonts}
\newcommand{\mgb}{MgB$_{2}$}
\newcommand{\tc}{\ensuremath{T_{\rm c}}}
\newcommand{\uphc}{$H_{\rm c2}$}

\begin{document}

\title{Specific heat of \mgb{} after irradiation}

\author{Yuxing Wang\dag,  Fr\'ed\'eric Bouquet\dag, Ilya Sheikin\dag, Pierre Toulemonde\dag, Bernard Revaz\dag, Michael Eisterer\ddag,
Harald W. Weber\ddag, Joerg Hinderer\S, Alain Junod\dag}

\address{\dag\ Universit\'e de Gen\`eve, D\'epartement de physique de la mati\`ere condens\'ee,}
\address{24 quai Ernest-Ansermet, CH-1211 Geneva 4, Switzerland}
\address{\ddag\ Atominstitut der \"Osterreichischen Universit\"aten, A-1020 Vienna, Austria}
\address{\S\ GHMFL, Max-Planck Institute Grenoble, BP 166, F-38042, Grenoble, France}

\begin{abstract}
We studied the effect of disorder on the superconducting
properties of  polycrystalline \mgb{} by specific-heat
measurements. In the pristine state, these measurements give a
bulk confirmation of the presence of two superconducting gaps with
$2 \Delta _0 /k{\rm_B} T_{\rm c} = 1.3$ and $3.9$ with nearly
equal weights. The scattering introduced by irradiation suppresses
\tc{} and tends to average the two gaps although less than
predicted by theory. We also found that by a suitable irradiation
process by fast neutrons, a substantial bulk increase of
$dH_{c2}/dT$ at \tc{} can be obtained without sacrificing more
than a few degrees in \tc. The upper critical field of the sample
after irradiation exceeds 28 T at T$\rightarrow 0$.

\end{abstract}

\pagebreak

\section{Introduction}
The recently discovered superconductor \mgb{} has attracted a lot
of attention, because of its unexpectedly high critical
temperature, \tc, for a phonon-mediated pairing
mechanism~\cite{twogap1,liu,choi,2,3,4}. It has been proposed that
such a high \tc{} is due to the existence of two superconducting
gaps~\cite{liu,choi}, with this claim supported by
experiments~\cite{3,4}. It was suggested that the two gaps open on
different parts of the Fermi surface~\cite{liu,choi}. One part is
three-dimensional, and arises from the $\pi$ bonding and
antibonding orbitals; it gives rise to the smaller gap with
$2\Delta_{\pi 0}/k_{\rm B}T_{\rm c}\sim 1.1 - 1.5$. The second
part consists of nearly cylindrical sheets, which arise from the
$\sigma$-band, and gives rise to the larger gap with
$2\Delta_{\sigma 0}/k_{\rm B}T_{\rm c}\sim 3.6 - 4.5$.

The anisotropy of the superconducting gaps and that of the Fermi
surface should lead to an anisotropic upper critical field, \uphc,
and this has indeed been reported in
\mgb~\cite{Hanisotropy,budko}. The anisotropy is temperature
dependent, the ratio of \uphc{} in the boron plane to that along
the $c$-axis being about 3 near \tc{} and about 7 at low
temperature.

An empirical two-band model based on the presence of two different
gaps was successfully used to fit specific-heat data obtained on
different \mgb{} samples in zero field~\cite{fred}. In this model,
each gap is assumed to follow the usual temperature dependence of
the BCS theory, whereas the gap ratio $2\Delta_0/k_{\rm B}T_{\rm
c}$ is allowed to differ from the BCS value 3.5. Both gaps are
expected to be sensitive to impurity scattering~\cite{liu}. An
interesting prediction is that non-magnetic interband scattering
will decrease \tc~ in the case of two coupled gaps, whereas
Anderson's theorem would rather predict that scattering is
irrelevant to first order for a single gap~\cite{golubov}. In the
limit of strong scattering, it has been predicted that both gaps
will be averaged up to the point where the single gap BCS limit is
recovered. This should occur for \tc~$\cong 27~$K~\cite{liu}. This
motivated the present study where the behavior of both gaps is
followed by bulk specific-heat experiments while disorder is
introduced in the material. We report heat capacity measurements
on a polycrystalline sample before and after irradiation by fast
neutrons. We found some suppression of the larger gap after
irradiation, whereas the smaller gap remains quite robust.

Finally, the normal-state resistivity and the irreversibility
field of \mgb{} were reported to be enhanced by proton
irradiation~\cite{caplin}. It has also been shown that neutron
irradiation can increase the upper critical field of this
material~\cite{eisterer}. Here we show that the upper critical
field of our sample increases from about 18 to 28~T after
irradiation. Compared to the results obtained on thin films by
oxygen post-annealing~\cite{larbal} or fast quenching~\cite{jap}
techniques, the specificity of this study lies in the use of a
bulk determination, in addition to magnetotransport measurements
performed up to 28~T in a Bitter-type magnet.

\section{Experimental details}

The polycrystalline sample (HP14) was synthesized from Mg (99.8\%)
and B (99.7\%) powders at 900$^\circ$C for 1 hour in a cubic press
at 3 GPa. Fig.~\ref{acdcsus}(a) shows the diamagnetic transition
of the pristine sample as detected by $ac$ susceptibility at 8 kHz
and 0.1 G. The midpoint of the transition is $\tc = 37~$K. Neutron
irradiation was subsequently performed at the Triga research
reactor in Vienna. The sample was irradiated to a fast neutron
fluence (E$>$0.1 MeV) of $10^{22}$ m$^{-2}$, and in a second
irradiation step a fluence of $2\times 10^{22}$ m$^{-2}$ was
added. Neutrons induce defects in \mgb~ mainly by neutron capture
of $^{10}$B followed by the emission of an alpha
particle~\cite{eisterer}. Due to the very small penetration range
of low energy neutrons in \mgb, thermal neutrons would induce
damage only at the surface of the sample, resulting in a highly
inhomogeneous defect structure. In order to avoid this problem, a
cadmium shield was used, which absorbs nearly all low energy
neutrons. The introduced disorder can be estimated to $1.7\times
10^{-2}$ and $ 5\times 10^{-2}$ dpa (displacements per atom) after
the first and the second irradiation, respectively.

The transition temperature was measured after each irradiation by
$ac$ susceptibility as shown in Fig.~\ref{acdcsus}. After
irradiation, \tc~shifted to lower temperature, 34.6~K after the
first and 30.2~K after the second irradiation, and the transition
broadened, but remained reasonably sharp compared to the result of
similar studies~\cite{caplin}. Therefore the disorder introduced
by neutron irradiation is fairly homogeneous.

The specific heat was measured before and after each irradiation
in the temperature range 2 -- 50 K by two different techniques. A
relaxation calorimeter was used from 2 to 16 K~\cite{4}, and an
adiabatic technique from 16 to 50 K~\cite{adiacalo}. The
resistivity of the sample was measured by the four-probe method.
Data at fields below 17 T were obtained in a superconducting
magnet by using a $dc$ current reversal technique, with a current
of 2~mA. Additional measurements from 12 to 28 T were performed in
a Bitter coil at the GHMFL in Grenoble using an $ac$ synchronous
detection technique. These measurements were performed with a
current of 2~mA at 11.7~Hz.

\section{Results and discussion}
\subsection{The increase of \uphc}

The specific heat of the sample was measured before and after each
irradiation in different magnetic fields. Fig.~\ref{sppic} shows
the specific heat difference $[C(B)-C(14{\rm T})]/T$. The baseline
$C$(14T) represents the normal state specific heat in the
temperature range shown in Fig.~\ref{sppic}; it was smoothed
before subtraction. Since the lattice contribution cancels after
subtraction, Fig.~\ref{sppic} in fact shows the difference between
the electronic specific heat in the superconducting and the normal
state.  The amplitude of the specific-heat jump in zero field
decreases after each irradiation. Its position is shifted in bulk
by the magnetic field to a lower temperature. The broadening of
the transition in increasing fields results from angular averaging
since our sample is polycrystalline and anisotropic. In \mgb, the
anisotropy of the \uphc{} is estimated as 2 -- 3 close to
\tc~\cite{4,Hanisotropy,budko}. Because of the transition
broadening due to anisotropy, the onset of the specific-heat
anomaly (see arrows in Fig.~\ref{sppic}) was used to define \tc.
This would correspond to the transition temperature of a single
crystal for $H$ parallel to the boron planes. At $H=0$, the
critical temperature is 37.4 K, 35.8 K and 31.2 K for the sample
before, after the first and after the second irradiation
respectively. Fig. \ref{hctc} shows the temperature dependence of
\uphc{} obtained from both specific-heat and resistance
measurements (before and after irradiation). Note that the
determinations based on the specific-heat jump are no longer
significant at high fields owing to the smearing of the jump.
Nevertheless, they unambiguously demonstrate that the average
initial slope $(dH_{\rm c2}/dT)_{T_{\rm c}}$ increases after each
irradiation.

A positive curvature (PC) of \uphc~near \tc~can been seen in Fig.
\ref{hctc} at different stages of irradiation. This can be
attributed to the existence of two superconducting gaps in the
sample, which may be considered as a particular case of gap
anisotropy~\cite{shulga,Schachinger}. It is pointed out by
theory~\cite{Schachinger,shulga2} that the PC disappears in the
``dirty" limit, where very strong interband scattering suppresses
the anisotropy. In order to investigate the change of PC after
irradiation, we used the following formula to fit our data near
\tc~up to 4 - 5 T: $H_{\rm c2}(T)=H_{\rm c2}^{\star}(1-T/T_{\rm
c})^{\alpha}$~\cite{Freuden}, where \tc~is the transition
temperature at zero field, $H_{\rm c2}^{\star}$ and $\alpha$ are
fitting parameter. We found $\alpha$ within the range of
$1.4\pm0.1$ before and after each irradiation, and no significant
trend was observed. This is consistent with the observation that
both gaps can still be identified after irradiation, confirming
that the interband scattering caused by disorder in this sample is
not strong enough to reach the ``dirty" limit, as discussed in
Section 3.2.

In order to verify the increase of \uphc{} after irradiation, we
performed resistance measurements in magnetic field up to 17~T on
the pristine sample and up to 28~T after the second irradiation.
The insets of Fig.~\ref{hctc} show typical field and temperature
sweeps after the second irradiation. For the pristine sample,
\uphc(0) is estimated to be about 18 T, in agreement with the
results obtained on single crystals for the field parallel to the
boron plane~\cite{Hanisotropy, fred2}. After the second
irradiation, the transition becomes broader at low temperature, as
it was already the case for the pristine sample, presumably again
as a consequence of anisotropy~\cite{Hanisotropy}. At $T=1.5$~K,
the transition is not complete even at 28 T, indicating \uphc(0)
higher than 28 T. Both specific-heat onsets and resistance
transition midpoints are plotted in the $H-T$ phase diagram of
Fig.~\ref{hctc}. The results of both determinations of $H_{\rm
c2}(T)$ agree reasonably well in the low field range, the onset of
calorimetric transitions appearing as an upper limit as expected.
The quasi-linear increase of $H_{\rm c2}(T)$ down to $T = 0$
differs remarkably from the horizontal slope at $T \rightarrow 0$
found for conventional superconductors. An explanation for this
unusual behavior was provided within the two band model
~\cite{shulga}. Note that a similar PC was found in several
borocabides~\cite{Freuden}, not necessarily implying the same
physical origin~\cite{shulga2}.

For a type-II superconductor in the dirty limit, $H_{\rm c2}(0)
\propto \gamma_n \rho T_{\rm c}$, where $\gamma_n$ is the
Sommerfeld coefficient in the normal state, and $\rho$ is the
normal-state resistivity at low temperature~\cite{orlando}. After
irradiation, \tc~decreases by 20$\%$, so that the main parameters
to be investigated for changes are $\gamma_n$ and $\rho$. As will
be shown later, $\gamma_n$ does not change significantly.

The residual resistivity depends sensitively on disorder. Before
irradiation, just above \tc, $\rho(45 \rm K)= 4.0~\mu\Omega.\rm
{cm}~$ and the residual resistance ratio is $RRR=\rho(300 \rm K) /
\rho(45 \rm K) = 3.3$. To estimate the electronic mean free path,
we use the following formula:

\begin{equation}\label{1}
l = [\frac{\pi k_{\rm B}}{e}]^2 \cdot \frac{\sigma}{\gamma_n
v_{\rm F}},
\end{equation}
where $\sigma$ is the normal-state conductivity and $v_{\rm F}$ is
the Fermi velocity. With
$\gamma_n=0.14~$mJ/K$^2$cm$^3$~\cite{junodnarlika} and $v_{\rm
F}=4.8\times 10^7~$cm/s~\cite{kortus}, this yields
$l\approx270~$\AA. Since $l\gg \langle\xi\rangle \approx
42~$\AA~\cite{note}, the sample is initially rather in the clean
limit. After the second irradiation, the low-temperature
resistivity increases to $\rho(45 \rm K)= 22.6 \mu\Omega.\rm
{cm}~$(RRR = 1.4), which gives $l\approx50~$\AA. The sample is now
in the intermediate case between the clean and dirty limit. In the
dirty limit, the upper critical field would be determined by the
geometrical average between the coherence length and the mean free
path $\xi=\sqrt{\xi_0 l}$. Therefore, at least qualitatively, one
is led to conclude that disorder induced by irradiation affects
\uphc~through the reduction of the mean free path. However, we
note that the sample approaches the dirty limit only after the
irradiation, so that the discussion given here remains essentially
qualitative. Moreover, the two-band nature of \mgb{} complicates
the determination of the mean free path~\cite{mazin2}.

Fig.~\ref{lowtsp} shows the specific heat below 16 K in different
magnetic fields at different stages of irradiation. At the highest
field accessible to our measurements, $\mu_0H=14$ T, $C/T$
reflects nearly the normal-state specific heat. At first view,
this would not seem to be the case, since $\mu_0$\uphc(0)~is 18
and 28 T before and after the second irradiation, respectively.
However, it was shown by several experiments that $\gamma$ versus
$H$ already saturates at about \uphc/2 in
polycrystals~\cite{4,note2}, so that no significant change is
expected above 14 T. This is also illustrated by the behavior of
the pristine sample (see Fig.~\ref{lowtsp}(a)), where specific
heat saturates near 8 T. In the inset of Fig.~\ref{lowtsp}, we
show that the low temperature specific heat at 14~T remains
unchanged at different stages of the irradiation, indicating that
the normal state Sommerfeld coefficient does not change (the
downturn that occurs at very low temperature in the $C/T$ versus
$T^2$ plot in the inset of Fig.~\ref{lowtsp} is believed to be due
to the presence of residual magnetic impurities). Therefore, as
already mentioned above, the variation of \uphc{} is not due to a
change in $\gamma_n$. We thus conclude that irradiation merely
enhances the scattering, causing both a decrease of \tc~ and an
increase of the upper critical field.

These results suggest that other methods able to introduce
disorder, such as fast quenching or chemical doping, could also
serve to increase \uphc, as was realized in A15
compounds~\cite{Flukiger}. Indeed, a large increase of \uphc~was
obtained by metallurgical heat treatment of \mgb~thin
films~\cite{larbal,jap}. The present study shows that this
increase of \uphc{} is a bulk property, at least at low fields,
where the specific-heat jump is clearly observable. It is
confirmed by direct resistance measurements at low temperatures
and high magnetic fields, rather than by extrapolation beyond 10 T
in previous work.

\subsection{Effect of irradiation on the gaps}

As discussed above, irradiation does not reduce the density of
states, but merely increases the scattering. For a one-band,
isotropic superconductor, scattering due to non-magnetic
impurities would account for the increase of the normal-state
resistivity, but would not affect \tc{} to first order. For a
two-band superconductor, $intra$band scattering would preserve the
two-gap structure even for a high concentration of
impurities~\cite{mazin2}. In the present case, a clear suppression
of \tc{} is observed; it can only be explained by $inter$band
scattering between the $\sigma$- and $\pi$-
bands~\cite{liu,golubov}. Therefore, a distinction should be made
between samples for which impurities affect only the normal-state
resistivity, leaving \tc{} unchanged~\cite{mazin2}, and the
present irradiated samples, for which both \tc{} and resistivity
are affected. These distinct situations reflect different balances
between intra- and interband scattering. If we believe that strong
interband scattering exists for the present samples, which is
rather unusual for \mgb~\cite{mazin2}, we expect changes in the
two-gap structure after strong irradiation. We proceed to study
this point experimentally.

The electronic specific heat $C_{\rm e}/T$ versus $T$ at zero
field is plotted in Fig.~\ref{shof2gaps}. The lattice contribution
determined at 14~T was smoothed and subtracted. The low
temperature excess of the specific heat with respect to a one-gap
BCS curve (hatched area in Fig.~\ref{shof2gaps}), which results
from the existence of the smaller gap~\cite{fred}, becomes less
pronounced after irradiation. A satisfactory fit can be obtained
for the unirradiated sample using a two-gap
model~\cite{fred,junodnarlika}. The fit gives $\Delta_0=6.1$ meV
for the larger gap and 2.1 meV for the smaller gap, with nearly
equal weights. Similar fits can be made after irradiation. The gap
values at each stage of irradiation are given in
Table~\ref{tabl1}. After the second irradiation, the quality of
the fit was degraded by the broadening of the transition. A
satisfactory fit was obtained by shifting \tc{} slightly above the
midpoint of the jump and by lowering $\gamma_n$ by 2.5\%.

An obvious effect of the irradiation is the suppression of the
larger gap, which, in a two-gap model, is reflected in the
specific heat jump at \tc~\cite{fred}. This gap, which originates
from the 2D $\sigma$-band, seems to be more sensitive to defects.
The smaller gap, which is reflected in the low temperature hump in
the specific heat near 10~K, appears to be quite robust in
absolute value, remaining within 10\% of the average 2.15 meV in
all cases. However, its reduced value $2\Delta_0/k_{\rm B}T_{\rm
c}$ increases after each irradiation, due to the decrease of \tc.
Liu et al. predicted that interband scattering in \mgb{} should
average both gaps, which would finally merge into a single BCS gap
when \tc~ reaches $\sim 27$ K~\cite{liu}. The observed gap values
are plotted as a function of \tc{} in the inset of
Fig.~\ref{shof2gaps}. This plot suggests that the single-gap limit
will only be reached for \tc{} smaller than the anticipated 27 K,
although additional irradiations will be required to prove it.
From the fits, one can also see that the weight of both gaps
remains almost unchanged after the irradiation.

Golubov et al. calculated the effect of interband scattering on
specific heat due to non-magnetic impurities within the two-band
model~\cite{golubov}, and pointed out that the impurity scattering
should increase the $\Delta C/ \gamma_{\rm n}T_{\rm c}$ ratio,
whereas the specific heat jump $\Delta C$ should remain nearly
constant. According to our experiment, $\Delta C$ decreases after
irradiation and $\Delta C/ \gamma_{\rm n}T_{\rm c}$ decreases only
slightly. This disagreement may be of experimental origin and due
to the transition broadening.

\begin{figure}
\caption{Specific heat $C/T$ at 3~K versus magnetic field on a
logarithmic scale, $\diamond~$before irradiation, $\bullet$ after
the first irradiation, $\bigtriangleup$ after the second
irradiation. Inset: from bottom to top, low temperature specific
heat at 0.5 T before irradiation, 0.84 T after the first
irradiation and 1 T after the second irradiation, respectively.}
\label{gammaH}
\end{figure}
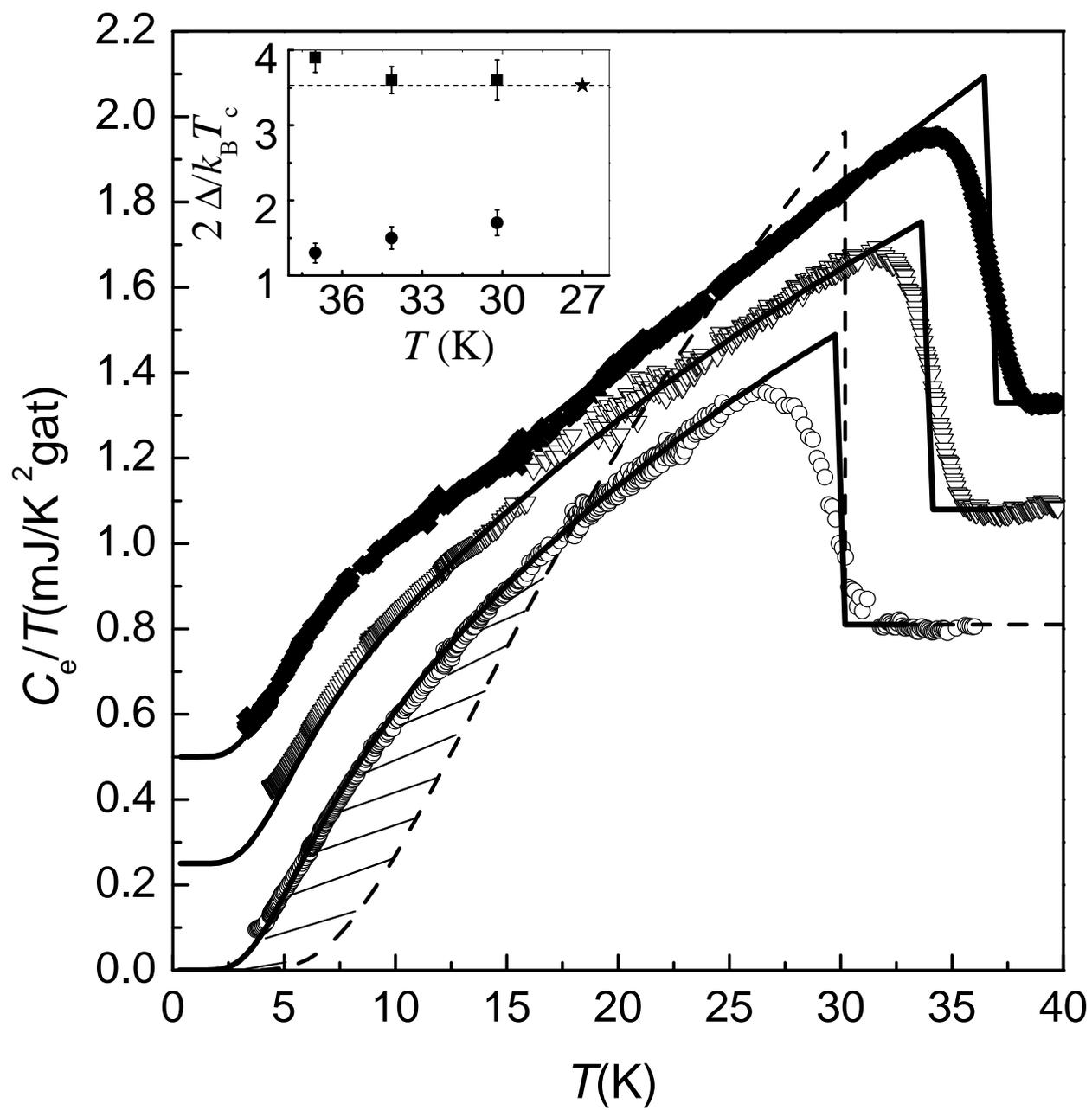
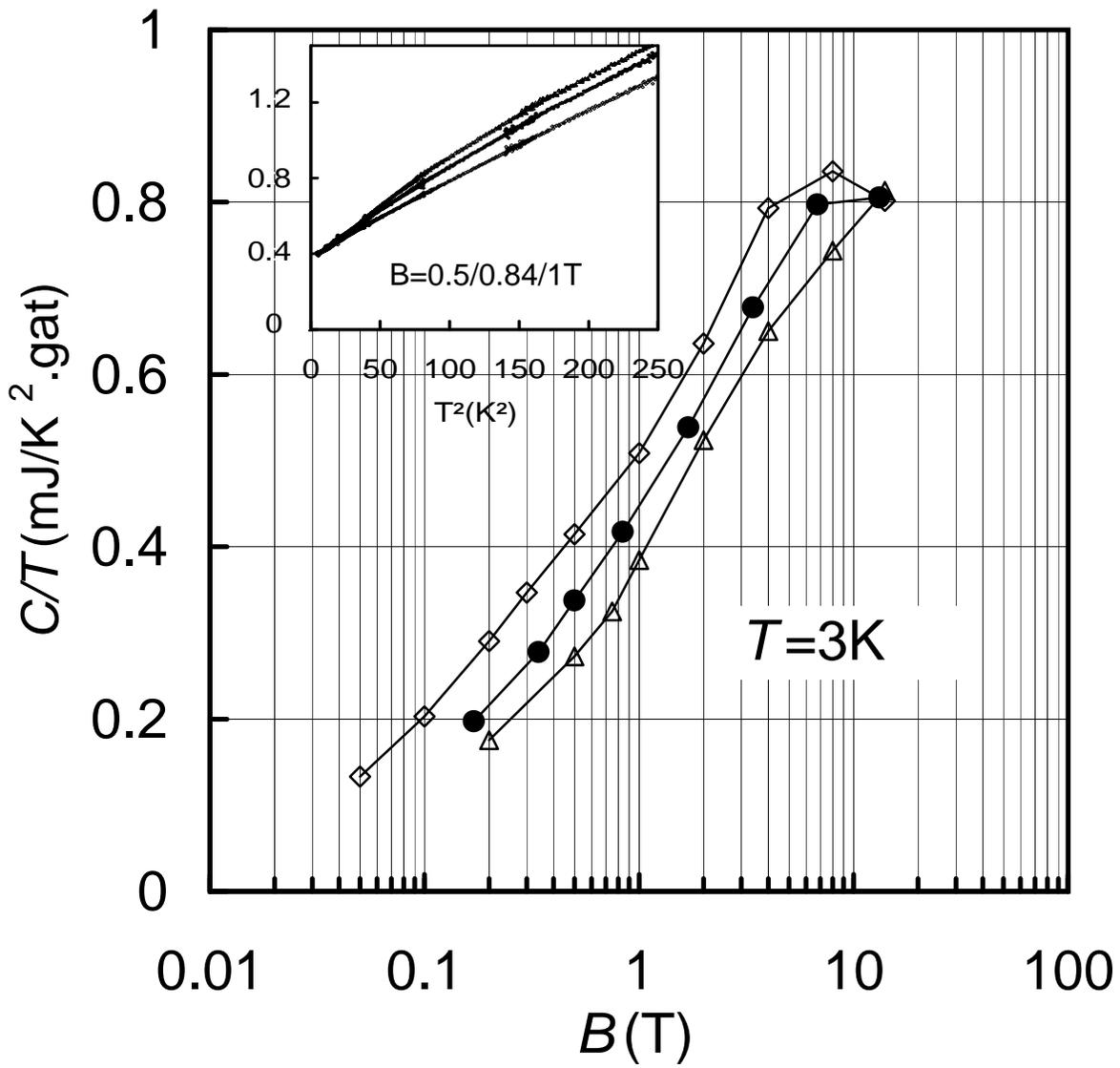

The coefficient of the mixed-state electronic specific-heat,
$\gamma(H)$, provides independent information on both gaps. Its
highly non-linear increase is shown by the plot $C(H,T)/T$ at
$T\ll T_{\rm c}$ (here, 3K) on a logarithmic field scale (see
Fig.~\ref{gammaH}). The curves at different stages of irradiation
lie nearly parallel to each other below 10 T, and saturate at the
same value of $\gamma_{\rm n}$ at 14 T. The small maximum at 8 T
for the non-irradiated sample is believed to be an artifact due to
a residual Schottky contribution of magnetic origin. The fact that
the three curves are parallel to each other implies a simple
relation $\gamma_{\rm pristine}(H)=\gamma_{\rm irradiated}(\alpha
H)$, where $\alpha\approx 1.7$ after the first irradiation and
$\alpha\approx 2$ after the second one. The inset of
Fig.~\ref{gammaH} illustrates this relation: the $C/T$ versus
$T^2$ curves at 0.5 T for the pristine sample, 0.84 T after the
first irradiation, and 1 T after the second irradiation point to
the same value of $\gamma(H, T=0)$.

According to theory~\cite{liu} and
experiment~\cite{fred,carrington}, both gaps have nearly equal
weights, i.e. the partial densities of states $N_{\sigma}$ and
$N_{\pi}$ of the associated bands, and the partial normal-state
Sommerfeld coefficients $\gamma_{\rm n \sigma}$ and $\gamma_{\rm n
\pi}$, are nearly equal. Recent scanning tunneling microscopy
measurements on a single crystal show that already in a magnetic
field of 0.2 T, there is a significant overlap of vortex cores in
the $\pi$-band vortices~\cite{eskildsen}. Very recent
specific-heat measurements on a single crystal have revealed that
$\gamma_{\pi}(H)$ increases very fast with the field, and
saturates above a crossover field $\mu_0 H \cong 0.4 \rm T$,
whereas $\gamma_{\sigma}(H)$ saturates at 3.3 T (field along the
$c$-axis) or 18~T (field in the $ab$-plane)~\cite{fred2}. It
follows that at low fields($B\leq 0.5~$T) the contribution from
the smaller gap dominates the behavior of $\gamma(H)$, in
agreement with theory~\cite{nakai}. On the contrary, at high
fields ($B>1~$T), the contribution from the smaller gap saturates,
so that the variation of $\gamma(H)$ comes from the larger gap.
The crossover near 0.4~T can be considered as the virtual upper
critical field $H_{\rm c2 \pi}$ for the smaller gap. The
parallelism of the curves in Fig.~\ref{gammaH} at high fields (1
-- 8 T) provides the proof for the increase of \uphc{} associated
with the larger gap. This is independently measured by the shift
of the specific-heat jump and resistive transitions.
Alternatively, at low fields ($B<0.5~$T), below the saturation of
the contribution $\gamma_{\pi}(H)$ associated with the smaller
gap, one may deduce from the approximate relationship
$\gamma_{\pi}(H)=\gamma_{\rm n \pi}H/H_{c2}$ that $H_{\rm c2 \pi}$
has increased after irradiation by the same factor $\alpha$ as
$H_{\rm c2 \sigma}$. This observation tends to support the
interpretation that the superconductivity in the $\pi$-band is not
independent, but rather induced by superconductivity in the
$\sigma$-band~\cite{nakai,eskildsen}. At this point we must recall
that our sample is a polycrystal, so that information on the
anisotropy is lost in the high field region dominated by the large
gap of the $\sigma$-band. More detailed information on this
unusual system would require irradiation experiments on single
crystals.

\section{Conclusion}
We studied irradiation effects on a polycrystalline sample of
\mgb. Irradiation by fast neutrons provides a good tool to
introduce $inter$band scattering. Specific-heat measurements show
a suppression of the larger gap after irradiation, whereas the
smaller gap $\Delta_{\pi}$ remains nearly unchanged. In terms of
reduced gap value $2\Delta_0/k_{\rm B}T_{\rm c}$, the gaps tend to
converge, but slower than predicted by considerations of the
interband scattering. The two-gap feature remains quite robust. We
also find that the upper critical or crossover field $H_{\rm c2
\pi}$ associated with the smaller gap follows the behavior of that
of the larger gap after irradiation. Together with resistivity
measurements, we show that a substantial increase of \uphc$~$ can
be obtained by irradiation, without sacrificing more than a few
degrees in \tc. If critical currents do not collapse, the
potential of \mgb~ tuned by disorder is worth further
investigations in view of high field superconducting applications.

\ack{This work was supported by the Swiss National Science
Foundation through the national Centre of Competence in Research
``Material with Novel Electronic Properties - MaNep''. The authors
thank R. Fl\"{u}kiger, J. Kortus, and A. Holmes for fruitful
discussions, and A. Lo, B. Seeber for technical assistance.}

\begin{table}[ht]
\begin{tabular}{cccc}
 \hline                 & Before irradiation    & After $1^{st}$ irradiation    & After $2^{nd}$ irradiation \\

\tc~(K)                 & 37           & 34.1         & 30.2\\

$\gamma_n~$(mJ/K$^2$.gat) & 0.83         & 0.83          & 0.81
\\

$\Delta_{\pi}$(meV)     & $2.07\pm0.20$         & $2.20\pm0.20$ &
$2.22\pm0.20$\\

$2\Delta_{\pi}/k_{\rm B}T_{\rm c}$  & 1.3&1.5&1.7\\

$\Delta_{\sigma}$(meV)  & $6.21\pm0.30$ & $5.30\pm0.25$ &
$4.68\pm0.35$\\

$2\Delta_{\sigma}/k_{\rm B}T_{\rm c}$ & 3.9 & 3.6 &3.6\\

$\gamma_{\rm n \pi}:\gamma_{\rm n \sigma}$    & 0.5:0.5      &
0.55:0.45 &0.55:0.45\\

$\Delta C$(mJ/K.gat)& $27.6\pm1$ & $25.3\pm1$ &$20.3\pm 1$\\

$\Delta C/\gamma_{\rm n} T_{\rm c}$   & 0.9 & 0.85 & 0.85\\ \hline
\end{tabular}
\centering \caption{Gap parameters of \mgb{} at different stages
of irradiation obtained by fits of the specific heat. $\gamma_n$
is the normal-state Sommerfeld coefficient. $\Delta_{\pi}$ and
$\Delta_{\sigma}$ are the gap values for the smaller and larger
gap. $\Delta C$ is the specific-heat jump. $\gamma_{\rm n \pi}$
and $\gamma_{\rm n \sigma}$ are the partial Sommerfeld
coefficients associated with the $\pi$- and $\sigma$-band,
respectively; $\gamma_{\rm n \pi} + \gamma_{\rm n \sigma} =
\gamma_{\rm n}$. } \label{tabl1}
\end{table}

\Figures
\begin{figure}
\caption{Diamagnetic transition of the sample measured by $ac$
susceptibility, real ($\chi'$) and imaginary ($\chi''$) part: (a)
before irradiation; (b) after the first irradiation; (c) after the
second irradiation.} \label{acdcsus}
\end{figure}

\begin{figure}
\caption{Transitions at \tc~in different magnetic fields measured
by specific heat, $\Delta C= C(B=0)-C(B=14 \rm T)$: (a) before
irradiation, from right to left $B=$~0, 0.5, 1, 2, 3 T, (b) after
the first irradiation, from right to left $B=~$0, 0.84, 1.7, 2.5,
3.4, 4.2, 5.1 T, (c) after the second irradiation, from right to
left $B=~$0, 0.5, 1, 2, 3, 4, 5, 6, 8 T. Note that the sample was
cut and its mass reduced for the irradiations, explaining the
increased scatter.} \label{sppic}
\end{figure}

\begin{figure}
\caption{Upper critical field \uphc$(T)~$ before and after
irradiation: $\triangle$ and full line, before the irradiation,
determined by the onset of the specific-heat jump; $\Box$, before
the irradiation, determined by the midpoint of the resistance
step, $\diamond$ and full line, after the first irradiation,
specific heat onset; $\circ$  and full line, after the second
irradiation, specific heat onset; $\bullet$, after the second
irradiation, transition midpoint in resistance. Inset: resistance
measurements after the second irradiation: upper right: magnetic
field sweeps at constant temperature, from left to right: $T=$~28,
24, 20, 16, 12, 8, 4, 1.5 K; lower left: temperature sweeps at
constant field, from left to right: $B=$~14, 12.5, 10, 8.5 T.}
\label{hctc}
\end{figure}

\begin{figure}
\caption{Total specific heat below 16~K in a $C/T$ versus $T^{\rm
2}$ plot: (a) before irradiation, from bottom to top: $B=~$0 ,
0.05, 0.1, 0.2, 0.3, 0.5, 1, 2, 4, 8, and 14~T; (b) after the
first irradiation, from bottom to top: $B=~$0, 0.17, 0.5, 0.84,
1.7, 3.4, 6.8, 14 T; (c) after the second irradiation, from bottom
to top: $B=~$0, 0.2, 0.5, 0.75, 1, 2, 4, 8 T. Inset: low
temperature specific heat at 14 T before and after each
irradiation (three data sets). The extrapolation to T = 0 shows
that $\gamma_{\rm n}$ does not vary.} \label{lowtsp}
\end{figure}

\begin{figure}
\caption{Electronic specific heat $C_{\rm e}/T$ versus $T$ with
two-gap fits\protect~\cite{fred}, from top to bottom: before
irradiation, after the first irradiation, after the second
irradiation, respectively. The latter two curves are shifted by
0.25 mJ/K$^{2}$gat for clarity. The dashed line represents the BCS
single-gap model with the same \tc{} and $\gamma_{\rm n}$ as the
sample after the second irradiation. Inset: variation of both gaps
as a function of \tc{} (note that the scale is reversed to show
increasing scattering from left to right). The dashed line is the
isotropic BCS limit. The star $\star$ is a calculated point where
the gaps should converge to the BCS value according to
Ref.~\protect\cite{liu}.} \label{shof2gaps}
\end{figure}

\end{document}